\documentclass[aps,prl,twocolumn,superscriptaddress,floatfix]{revtex4-2}

\usepackage{amssymb}
\usepackage{amsmath}
\usepackage{graphicx}
\usepackage{dcolumn}
\usepackage{bm}
\usepackage{amsmath}
\usepackage{soul}
\usepackage{xcolor}
\usepackage[dvipsnames]{xcolor}
\usepackage[colorlinks=true,allcolors=blue]{hyperref}

\DeclareRobustCommand{\scottl}[1]{ \colorbox{LimeGreen!30}{(scott:) \href{#1}{link}} }

\newcommand{\hata}{\hat{a}}
\newcommand{\hatb}{\hat{b}}

\newcommand{\hath}{\hat{H}}
\newcommand{\hatn}{\hat{n}}

\newcommand{\hathm}[1]{\hat{H}_\mathrm{#1}}
\newcommand{\lr}[1]{\langle{#1}\rangle}

\begin{document}

\title{Nanophotonic control of collective many-body states in Kerr solitons}

\author{Yan Jin}
\email{yan.jin@colorado.edu}
\affiliation{Time and Frequency Division, National Institute of Standards and Technology, Boulder, Colorado 80305, USA}
\affiliation{Department of Physics, University of Colorado Boulder, Boulder, Colorado 80309, USA}

\author{Jizhao Zang}
\affiliation{Time and Frequency Division, National Institute of Standards and Technology, Boulder, Colorado 80305, USA}
\author{Sarang Yeola}
\affiliation{Time and Frequency Division, National Institute of Standards and Technology, Boulder, Colorado 80305, USA}
\affiliation{Department of Physics, University of Colorado Boulder, Boulder, Colorado 80309, USA}

\author{Alexa R. Carollo}
\affiliation{Time and Frequency Division, National Institute of Standards and Technology, Boulder, Colorado 80305, USA}

\author{Nitesh Chauhan}
\affiliation{Time and Frequency Division, National Institute of Standards and Technology, Boulder, Colorado 80305, USA}
\affiliation{Department of Physics, University of Colorado Boulder, Boulder, Colorado 80309, USA}

\author{Scott B. Papp}
\email{scott.papp@nist.gov}
\affiliation{Time and Frequency Division, National Institute of Standards and Technology, Boulder, Colorado 80305, USA}
\affiliation{Department of Physics, University of Colorado Boulder, Boulder, Colorado 80309, USA}

\date{\today}

\begin{abstract}
Spatially periodic systems of coupled bosons are governed by on-site interactions and tunneling between sites, opening a rich phase space of many-body behavior. Here, we explore nanophotonic control of collective many-body light states in a driven-dissipative Kerr microresonator. We demonstrate a non-equilibrium Mott insulator to superfluid transition that arises from the interplay of spatially local Kerr interactions that generate and mediate interference among discrete frequency modes. A photonic-crystal (PhC) lattice bandgap inscribed on the resonator controls linear mode coupling while preserving self-mode Kerr interactions. By increasing the PhC bandgap, we suppress nonlinear cross-mode coupling to access the Mott-insulator phase, wherein the soliton spectrum forms a flattop frequency comb with large and uniform power per mode. In contrast, reducing the PhC bandgap restores cross-mode coupling and drives a delocalized superfluid regime characterized by long-range phase coherence and a spectrum with non-uniform power distribution. Our work shows that many-body physics creates collective states in driven-dissipative systems, enabling advances in programmable photonics and quantum-optical computing.

\end{abstract}

\maketitle

Interacting bosons in a periodic potential exhibit rich many-body behavior arising from the competition between kinetic delocalization and on-site repulsion \cite{BlochDalibardZwerger2008}. The framework for this physics is the Bose--Hubbard Hamiltonian $\hat H = -J\sum_{i,j}\hat a_i^{\dagger}\hat a_j
+ \tfrac{U}{2}\sum_i \hat n_i(\hat n_i-1)
- \mu_\mathrm{BH}\sum_i \hat n_i$, where $J$ represents tunneling between neighboring sites $i$ and $j$, $U$ is the on-site interaction energy, $\mu_\mathrm{BH}$ is the chemical potential, $\hata_i$ and $\hata_i^\dagger$ are bosonic anhilation and creation operators, and $\hatn_i=\hata_i^\dagger \hata_i$.  When tunneling dominates, particles are delocalized across the lattice and form a phase-coherent superfluid.  When on-site interactions dominate, the system enters a Mott insulator characterized by integer number occupancy per site and loss of long-range phase coherence.  The transition between these regimes constitutes a paradigmatic example of a quantum phase transition in correlated bosonic matter, which has been explored in ultracold atoms \cite{Greiner2002, bakr2010probing, tomita2017observation, weckesser2025realization}, superconducting circuits \cite{Ma2019},  Josephson junctions \cite{Fazio2001, Houck2012}, exciton-polaritons \cite{Lerario2018}, and photons \cite{Hartmann2007}.

Kerr solitons are discrete, nonlinear excitations of an open microresonator field, sustained by a coherent external drive and the spatially local Kerr interaction \cite{Lugiato1987, leo_temporal_2010, herr2014temporal, cole_soliton_2017}. Outcoupled solitons form a periodic ultrafast pulse train, which we associate with a frequency comb \cite{Spencer2018, drake_terahertz-rate_2019, drake_thermal_2020}. The mean-field depiction of the bosonic resonator comb ($\hata$) with mode numbers $\mu$ comes from the quantization $[\hata_{\mu},\hata_{\mu^\prime}^\dagger]=\delta_{\mu\mu^\prime}$ and the dynamics of the quantized Lugiato-Lefever equation (LLE) $\dot{\hat{a}}_\mu = \frac{1}{i\hbar}[\hat{a}_\mu, \hath_\mathrm{tot}] - \hat{a}_\mu$ for all contributions to the Hamiltonian \cite{PhysRevA.93.033820}. Focusing on the Kerr interaction highlights the resonator coordinate $\theta$ and the spatial and mode basis representations of $\hat H_{\mathrm{Kerr}}=\frac{\hbar}{2}\int d\theta \, \hata^\dagger(\theta)\hata^\dagger(\theta)\hata(\theta)\hata(\theta) = \frac{\hbar}{2}\sum_{\mu_1,\mu_2,\mu_3,\mu_4}\delta_{\mu_1+\mu_2,\mu_3+\mu_4}\,\hat a_{\mu_1}^\dagger\hat a_{\mu_2}^\dagger\hat a_{\mu_3}\hat a_{\mu_4}$, respectively, where $\delta$ is the Kronecker delta function. The pump laser power and detuning and particularly mode dispersion simultaneously interact with $\hat H_{\mathrm{Kerr}}$, making these powerful control parameters to explore soliton behavior. The ability to harness many-body Kerr interactions will benefit advanced microcomb applications that demand high capacity, such as data communications \cite{pirmoradi_integrated_2025}, microwave photonics \cite{zang_universal_2025}, and photonic information processing \cite{jin_kerr_2025}. 

Coherent backscattering in photonic-crystal microresonators (PhCRs) has been used to control dispersion through a photonic bandgap (BG), resulting in advanced soliton microcomb capabilities \cite{Yu2020, Yu2021, Black:22, Lucas2023TailoringMicrocombs, Zang_2025, jin2025bandgap, ulanov2025quadrature, ulanov2024synthetic, Liu:25}. PhCR devices incorporate a nanoscale modulation with a periodicity keyed to a specific azimuthal mode, inducing a frequency splitting set by the backscattering rate. Therefore, each mode has forward (${\hat{a}}_\mu$) and backward (${\hat{b}}_\mu$) propagating components, described by a coupled-mode LLE, where the total Hamiltonian sums to $\hath_\mathrm{tot} = \hath_\mathrm{free} + \hath_\mathrm{pump} + \hath_\mathrm{Kerr} + \hath_\mathrm{BG}$. The resonator term $\hathm{free}$ involves $\alpha+\frac{1}{2}d_2\mu^2(\hata_\mu^\dagger\hata_\mu+\hatb_\mu^\dagger\hatb_\mu)$, where $\alpha$ is the pump detuning and $d_2\mu^2/2$ is the normalized second-order dispersion; the $\hath_\mathrm{pump}$ term is $i\hbar f(\hata_0^\dagger-\hata_0)$ with pump amplitude $f$; the $\hath_\mathrm{BG}$ term is $-\hbar \sum_\mu (\gamma_\mu \hata_\mu^\dagger \hatb_\mu + \gamma_\mu^* \hatb_\mu^\dagger \hata_\mu)$ with half-bandgap $\gamma_\mu$; and the four-mode Kerr interaction $\hath_\mathrm{Kerr}$ contains a series of terms $-\frac{\hbar}{2} (\hata_n^\dagger \hata_q^\dagger \hata_m \hata_p + \hatb_n^\dagger \hatb_q^\dagger \hatb_m \hatb_p)$ with four summation indices \cite{Kondratiev2020}. In the mean-field regime, the driven-dissipative bosonic comb modes $\psi_\mu = (a_\mu, b_\mu)$ with $a_\mu = \lr{\hata_\mu}$ and $b_\mu = \lr{\hatb_\mu}$ constitute a collective many-body state of light.

\begin{figure*}[t!] \centering%
\includegraphics[width=0.95\textwidth]{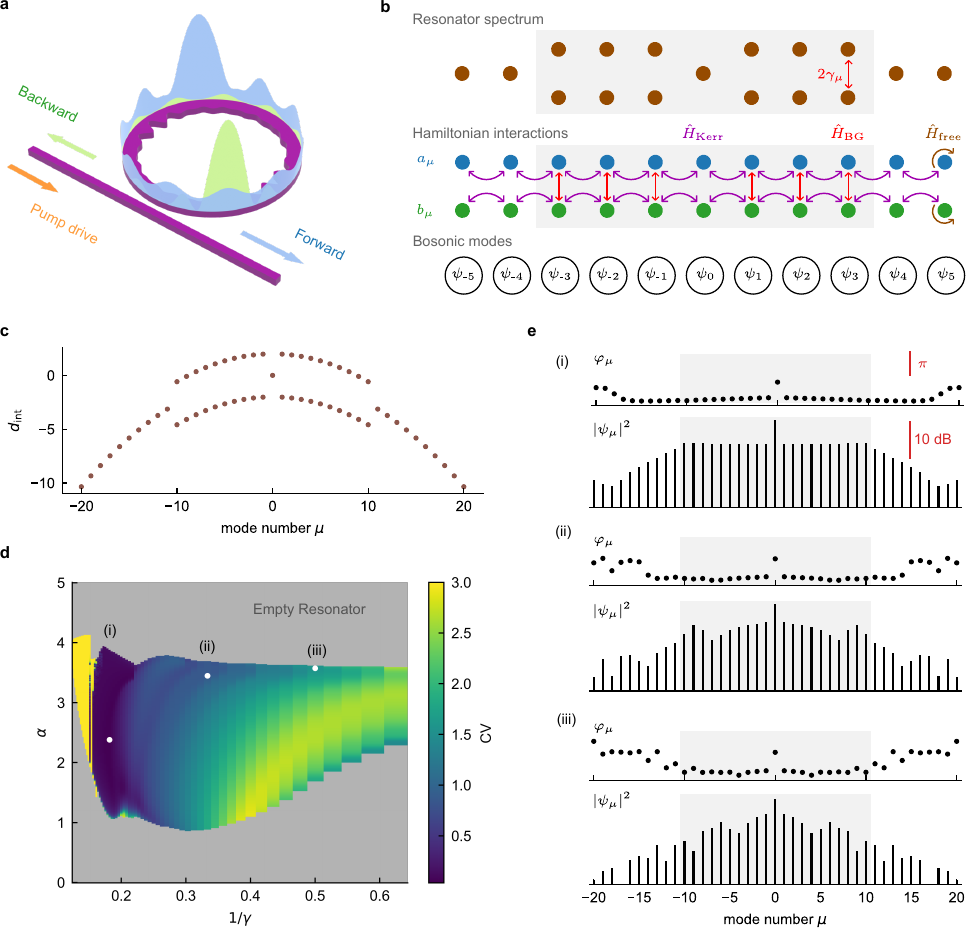}%
\caption{Mott insulator to superfluid transition in a Kerr soliton with a lattice bandgap. \textbf{a}, Schematic of forward (blue) and backward (green) two-component soliton on a PhCR excited by the pump (orange). \textbf{b}, The lattice of $\psi_\mu$ in the mode domain. First row (brown markers): resonator modes occupying each site of the lattice. The split modes have a normalized bandgap $2\gamma_\mu$. Second row: the interplay of $\hathm{free}$, $\hathm{Kerr}$ and $\hathm{BG}$ across the lattice. Third row: lattice site for $\psi_\mu$. \textbf{c}, Normalized $d_\mathrm{int}(\mu)$ with $\mu_s\in\{\pm1, \pm2, ...\pm10\}$. \textbf{d}, The CV map with respect to $\alpha$ and $1/\gamma$. Three representative cases (i)(ii)(iii) with different $\gamma$ for \textbf{e}. \textbf{e}, Mott insulator to superfluid transition presented by the phase and spectra of $\psi_\mu$ from (i), (ii) to (iii). }.  \label{Fig1}
\end{figure*}

Here, we explore a non-equilibrium Mott-insulator to superfluid transition in driven-dissipative Kerr microresonator solitons, using photonic bandgaps to control cross-mode interactions and interference across the comb. We employ a PhCR with $N$ engineered modes, such that $\hat H_{\mathrm{BG}}$ arises directly from the nanostructured geometry. This coupling in the resonator mode index tunes the competition between self- and cross-interactions through $\hat H_{\mathrm{Kerr}}$. Enabled by universal phase matching for Kerr nonlinear processes, we generate soliton microcombs across the Mott-insulator to superfluid phase space by exciting selected modes of $\hat H_{\mathrm{BG}}$ with a continuous-wave (CW) pump. Optical spectra and photodetection reveal two distinct regimes of the bosonic fields: a flattop comb with nearly uniform mode occupation and a strongly modulated comb arising from cross-mode Kerr interactions. Importantly, controllable access to the Mott-insulator and superfluid regimes does not rely on a single BG configuration. We demonstrate access to the full phase diagram both with a lattice BG in which $\gamma_0=0$ for the pumped mode and with PhCRs in which only the pump mode has a BG. Our work shows that engineered photonic BGs transform Kerr microresonators into many-body photonic systems, where collective phases of light can be designed, accessed, and controlled.

Figure~\ref{Fig1}a illustrates the driven–dissipative Kerr PhCR, where modulation of the wall imposes a lattice BG and a waveguide provides external coupling. We control $\hat H_{\mathrm{BG}}$ by superposing a periodic variation of the inner wall radius, $\rho_\mathrm{in}=\rho + \sum_{\mu\in\mu_s}\rho_\mu\sin\!\big(2(m_0+\mu)\theta\big)$, where $\rho$ is the inner wall radius, $\rho_\mu$ sets the contribution, $2\,m_0$ defines the modulation period for the azimuthal mode of the pump laser, and $\mu_s$ includes the modes with nonzero $\gamma_\mu$.  The BG couples forward- (blue) and backward-propagating (green) modes and their field amplitudes, which, under CW pumping (orange), generate counter-propagating soliton fields. Figure~\ref{Fig1}b presents a schematic of the PhCR mode structure, highlighting the lattice BG, the unsplit pump mode, and the connection to the bosonic comb mode fields $\psi_\mu$. The brown markers in the first row denote the resonator modes occupying each site of the lattice. Modes subject to the PhC acquire a normalized splitting of $2\gamma_\mu$. The second row of Fig.~\ref{Fig1}(b) summarizes the dominant Hamiltonian contributions to the resonator dynamics, and the bottom row indicates the bosonic modes made up of $\psi_\mu = (a_\mu, b_\mu)$. To show how the settings of $\gamma_\mu$ affect the dynamics of $\psi_\mu$, we analyze an example of a PhCR operating in the normal-dispersion regime ($d_2<0$) with a 20 mode lattice BG, $\mu_s = \{\pm1, \pm2, ..., \pm10\}$. As shown in Fig.~\ref{Fig1}c, the normalized integrated dispersion follows $d_\mathrm{int}(\mu) = \frac{1}{2} d_2 \mu^2 \pm \gamma_\mu$, where $\gamma_\mu = \gamma$ for $\mu \in \mu_s$ and zero otherwise.

Using the Hamiltonians in Fig.~\ref{Fig1}b and ~\ref{Fig1}c, we compute the mean-field phase diagram by the coupled-mode LLE equations (Methods), where $\psi_\mu$ denotes the steady-state modal amplitudes obtained for experimentally controlled parameters $\alpha$ and $1/\gamma$; see Fig.~\ref{Fig1}d. In this driven–dissipative Kerr system, the key observable distinguishing the phases is spectral flatness, which we use to characterize the many-body light states. We compute the coefficient of variation (CV), defined as the standard deviation normalized by the mean modal intensity over the lattice BG,
\begin{equation}
    \mathrm{CV} = \frac{\mathrm{Std}_{\mu \in \mu_s}(|\psi_\mu|^2)}{\mathrm{Mean}_{\mu \in \mu_s}(|\psi_\mu|^2)}.
\end{equation}
A small CV indicates uniform occupation consistent with the Mott insulator, whereas a large CV reflects enhanced spectral interference associated with the superfluid regime. To construct the phase diagram, we sweep $\gamma$ and $\alpha$ in an LLE calculation and identify all non-trivial steady-state solutions. As $\gamma$ increases (i.e., as $1/\gamma$ decreases), the CV decreases, marking the transition from the superfluid to the Mott-insulator state. In the high-CV region, only a signal-idler pair is excited, while the remaining split modes remain unoccupied, reflecting the inability to phase match across the full lattice BG.

To visualize the evolution across the phase diagram, we select three representative cases from Fig.~\ref{Fig1}d: (i) $\gamma = 5.5$, (ii) $\gamma = 3$, and (iii) $\gamma = 2$, and plot the corresponding relative phase $\varphi_\mu = \arg(a_\mu/b_\mu)$ together with the total spectra $|\psi_\mu|^2$ in Fig.~\ref{Fig1}e. The spectrum quantifies amplitude uniformity with a clear plateau in the Mott-insulator regime (i) and large mode-to-mode variation in the superfluid regime (iii). The phase profiles show the stiffness of the underlying many-body state: a phase that varies smoothly across $\mu$ indicates a collectively ordered state, whereas a fragmented phase profile reflects enhanced mode competition and reduced phase stiffness.


\begin{figure*}[t!] \centering%
\includegraphics[width=1\textwidth]{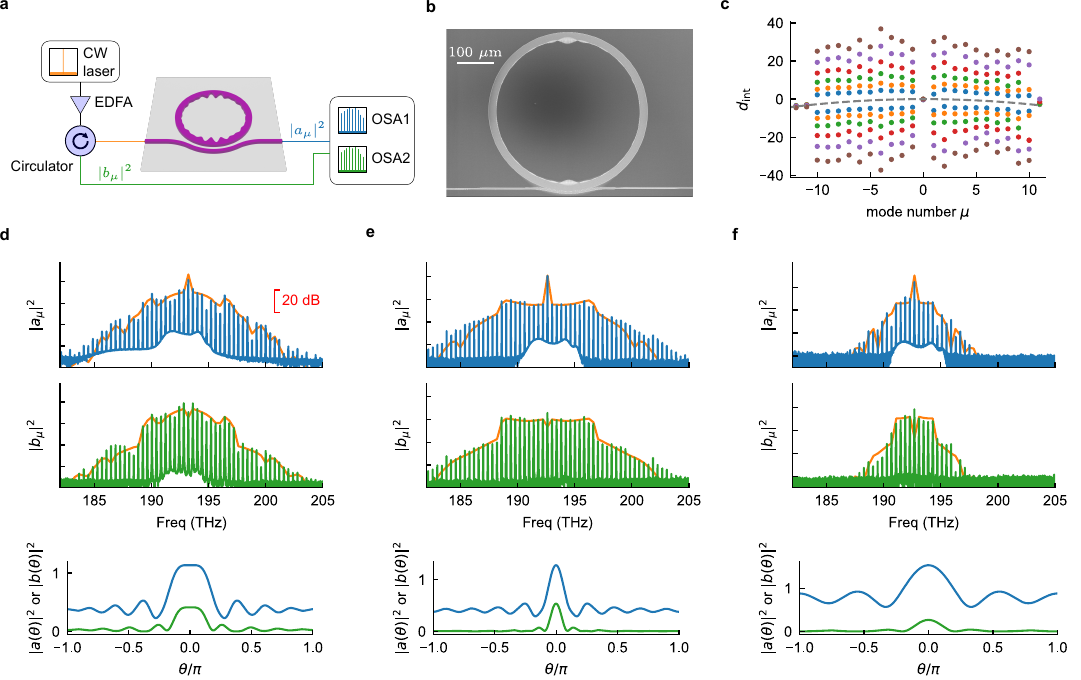}%
\caption{Observation of Mott-insulator and superfluid states in Kerr solitons. \textbf{a}, Experimental setup. CW: continuous wave. EDFA: erbium-doped fiber amplifier. OSA: optical spectrum analyzer. \textbf{b}, SEM image of a lattice-BG PhCR, exaggerated to show the PhC pattern. \textbf{c}, Normalized dispersion of 6 devices (marked in different colors) with increasing $\gamma_\mu$. \textbf{d-f}, Forward (blue) and backward (green) spectra and pulses. The orange curves are the simulated spectra. Every division is 20 dB for the spectra axis. The pulses $|a(\theta)|^2$ or $|b(\theta)|^2$ are normalized to $|f|^2$. \textbf{d}, Superfluid state with a 20 mode lattice BG. \textbf{e}, Mott insulator state with a 20 mode lattice BG. \textbf{f}, The Mott insulator state with an 8 mode lattice BG.
}\label{Fig2}
\end{figure*}

We experimentally access the Mott-insulator and superfluid regimes in PhCR devices engineered to implement a lattice BG; see Fig.~\ref{Fig2}. Operationally, we fabricate a series of devices to systematically tune $\gamma$. By generating soliton microcombs in these PhCRs and directly measuring the forward and backward spectra, we observe controlled access to flattop and strongly modulated comb states. Figure~\ref{Fig2}a shows the setup for our experiments. A continuous-wave (CW) laser in the 1550 nm band provides the pump, which is amplified by an erbium-doped fiber amplifier (EDFA) and coupled into a tantalum pentoxide ($\text{Ta}_2\text{O}_5$, hereafter tantala) PhCR. An optical circulator separates out the pump, and we have immediate access to the forward port, enabling independent spectral measurement of $|a_\mu|^2$ and $|b_\mu|^2$. We measure $|a_\mu|^2$ and $|b_\mu|^2$ with optical spectrum analyzers (OSA1 and OSA2).

Figure~\ref{Fig2}b shows a scanning electron microscope (SEM) image of a fabricated PhCR, illustrating the superimposed nanostructure of $\rho_{\mu}$ that generates the lattice BG. To demonstrate systematic control of $\gamma$ with the lattice BG, we fabricate six devices with identical radius and ring waveguide width but with varying $\rho_\mu$. We characterize the resulting lattice BG through measurements of $d_\mathrm{int}(\mu)$, as shown in Fig.~\ref{Fig2}c. To extract $d_\mathrm{int}(\mu)$, we scan a widely tunable CW laser across the PhCR resonances and record the transmission spectrum. The resonance frequencies are identified for each azimuthal mode, allowing reconstruction of the lattice-BG spectrum. In the presence of PhC modulation, the forward and backward modes hybridize, producing a resolvable splitting in the resonances for $\mu \in \mu_s$. The dashed curve in Fig.~\ref{Fig2}c indicates the fitted average quadratic dispersion, confirming the consistent background normal dispersion across devices while highlighting the engineered mode splittings introduced by the lattice BG. Especially for larger settings of $\rho_{\mu}$, we observe variability in the nanofabricated PhCR structure that appears as BG variation. 

\begin{figure*}[t!] \centering%
\includegraphics[width=1\textwidth]{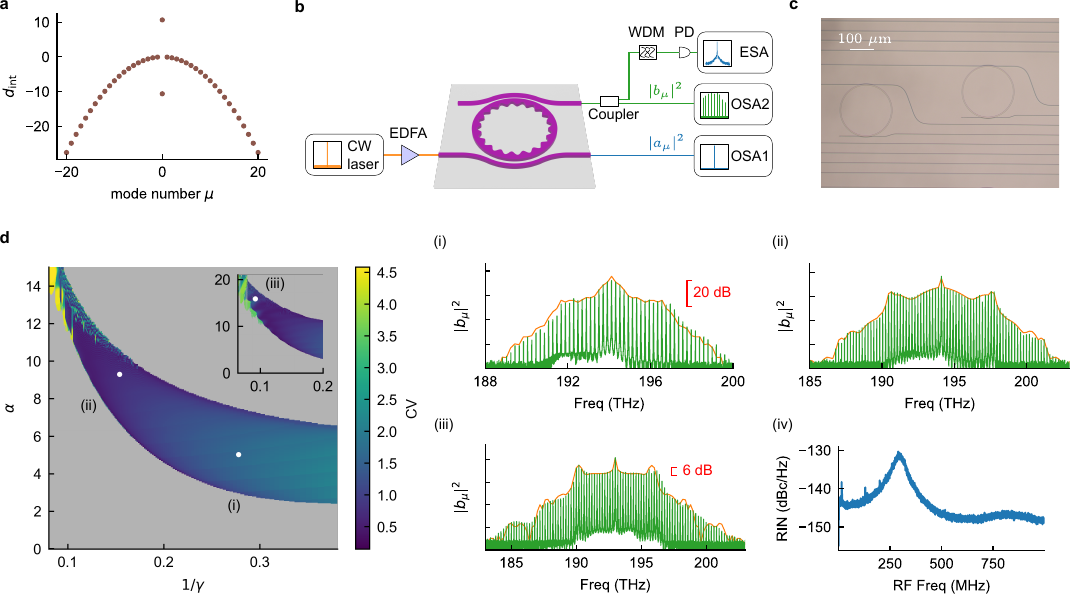}%
\caption{Mott-insulator to superfluid regimes in the pump-BG case. \textbf{a}, Integrated dispersion of the pump-BG. \textbf{b}, Setup for the PhCRs, including drop ports to directly measure forward and backward fields. WDM: wavelength division multiplexer. PD, photodiode. ESA, electrical spectrum analyzer. \textbf{c}, Microscopic image for devices. \textbf{d}, CV map with $f=2.8$ and $d_2=-0.05$ calculated over $\mu_r$ in the pump-BG regime. The inset is calculated with $f=4.2$ and $d_2=-0.12$. Spectra in (i)(ii)(iii) correspond to the white dots in the CV map, and the RIN is presented in (iv).
}\label{Fig3}
\end{figure*}

Figure~\ref{Fig2}d--f present experimental signatures of Mott-insulator and superfluid behavior amongst the bosonic comb modes. To understand the experiments, we write a decomposition of the relevant Hamiltonian terms $\hat H_{\mathrm{self}} = \hathm{free} + \hat H_{\mathrm{BG}} + \hat H_{\mathrm{SPM}}$ and $\hat H_{\mathrm{cross}} = \hat H_{\mathrm{CPM}} + \hat H_{\mathrm{FWM}}$. Here $\hat H_{\mathrm{SPM}}$ denotes the self-phase modulation contribution, while $\hat H_{\mathrm{CPM}}$ and $\hat H_{\mathrm{FWM}}$ contain cross-phase modulation and four-wave mixing processes, respectively; see Methods. Hence, $\hat H_{\mathrm{BG}}$ increases the effective self-interaction scale, shifting the balance between $\hat H_{\mathrm{self}}$ and $\hat H_{\mathrm{cross}}$ while leaving other terms intact. For a lattice BG of $N=20$ and a relatively small $\gamma = 3.7$, cross-mode interactions dominate over BG-induced hybridization and the spectra exhibit strong modulation across $\mu$; see Fig.~\ref{Fig2}d. The panel shows experimental measurements of $|a_\mu|^2$ (blue), corresponding to the forward-propagating mode family, and $|b_\mu|^2$ (green), corresponding to the backward-propagating mode family, plotted as a function of $\mu$. We measure the forward and backward CV as 1.2 and 1.7, respectively. The orange curves represent spectra predicted from the Hamiltonian model, which are subsequently used to reconstruct the corresponding intracavity pulse shapes in the time domain. The non-uniform occupation across the bosonic comb is consistent with a superfluid in which inter-mode coherence mediated by $\hat H_{\mathrm{CPM}}$ and $\hat H_{\mathrm{FWM}}$ remains effective.

As the BG is increased to $\gamma = 5.2$ (Fig.~\ref{Fig2}e), the contribution of $\hat H_{\mathrm{BG}}$ grows and the balance shifts toward $\hat H_{\mathrm{self}}$. In this regime, the spectra in both directions collapse into a pronounced plateau, forming a flattop soliton microcomb. The forward and backward mode families become rigidly hybridized, and the occupation of each mode becomes increasingly governed by local interactions rather than inter-mode exchange. Consequently, the forward and backward CV are reduced to 0.4. The reconstructed time-domain pulses narrow accordingly, reflecting the enhanced spectral uniformity and phase rigidity characteristic of the Mott-insulator regime. The agreement between measured spectra and Hamiltonian-based reconstructions confirms that tuning $\gamma$ directly controls the competition between $\hat H_{\mathrm{self}}$ and $\hat H_{\mathrm{cross}}$, reshaping the collective state of the bosonic comb modes.

To further test the generality of this Hamiltonian competition, we examine a third device type with a reduced lattice consisting of eight strongly split modes, $\mu_s = \{\pm1, \pm2, \pm3, \pm4\}$, shown in Fig.~\ref{Fig2}f. Despite the smaller lattice BG size, the same balance between $\hat H_{\mathrm{self}}$ and $\hat H_{\mathrm{cross}}$ governs the observed behavior. As the BG is increased, $\hat H_{\mathrm{BG}}$ again enhances the effective self-interaction, and the spectra evolve toward a flattop distribution, accompanied by the corresponding narrowing of the reconstructed time-domain pulses.

To explore the universality of the Mott-insulator to superfluid transition in soliton microcombs, we fabricate PhCRs with a fundamentally different BG design; see Fig.~\ref{Fig3}. In contrast to the lattice BG, here we engineer a single resonance splitting only at the pump mode by introducing one periodic modulation with $\gamma_\mu = \gamma\,\delta_{\mu,0}$. This pump BG isolates the interplay between $\hat H_{\mathrm{BG}}$, $\hathm{free}$, and $\hat H_{\mathrm{Kerr}}$, while preserving the same competition between $\hat H_{\mathrm{self}}$ and $\hat H_{\mathrm{cross}}$. For phase-matching in the pump-BG case, we excite the lower frequency split-mode branch, and under Kerr dynamics the backward pump occupation $|b_0|^2$ increases, setting the nonlinear interaction scale of all comb modes. Moreover, as $\gamma$ increases, the pump detuning $\alpha$ must also increase to sustain a soliton, enabling a direct adjustment of $\hathm{free}$ that increases the interaction scale and the per-mode power of the bosonic comb. In this way, $\gamma$ continues to function as a control parameter that strengthens $\hathm{self}$ relative to $\hathm{cross}$, driving the system toward the Mott-insulator regime.

Figure~\ref{Fig3}a shows the $d_\mathrm{int}$ profile for a normal dispersion pump-BG PhCR. There is no extended spectral feature that isolates a block of modes, nevertheless, dispersion-induced walk-off naturally limits phase-matched four-wave mixing. As a result, a central band of comb modes, which we denote $\mu_r$, become stabilized while outer modes form dispersive waves, providing a dynamical mechanism that fixes the comb bandwidth. An approximation of $\mu_r$ is $\{\pm1, \pm2, ... \pm \lfloor\sqrt{\frac{-2\alpha}{d_2}}-1\rfloor \}$, where $\lfloor \cdot \rfloor$ is the floor function; see Methods. We use PhCR devices fabricated with a drop port, enabling measurement of $|b_\mu|^2$ from a separate output as the pump laser. The measurement setup (Fig.~\ref{Fig3}b) incorporates a coupler after the drop port for relative intensity noise (RIN) measurements. A microscope image of a representative device is shown in Fig.~\ref{Fig3}c.

\begin{figure*}[t!] \centering%
\includegraphics[width=1\textwidth]{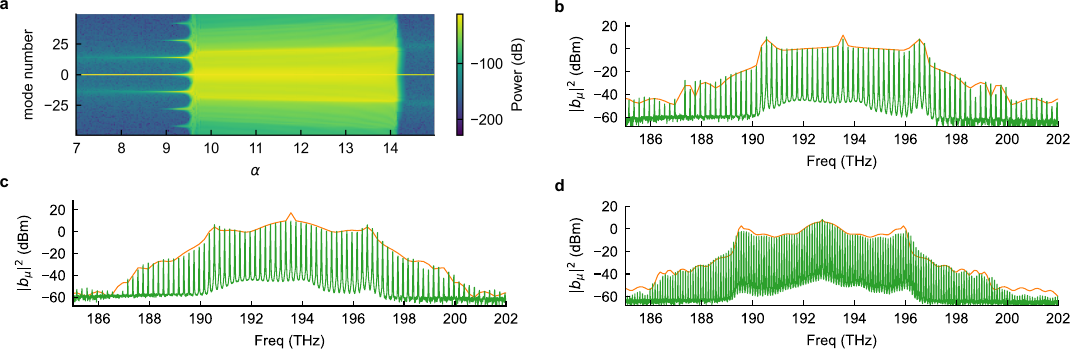}%
\caption{High-power pump-BG soliton microcombs. \textbf{a}, Soliton microcomb spectrum as $\alpha$ is tuned, illustrating the transition into stable high-power states. \textbf{b}, Mott-insulator regime at 200 GHz repetition rate, exhibiting a flattop spectral envelope with high power and reduced spectral variance. \textbf{c}, Superfluid regime at 200 GHz repetition rate, characterized by spectral modulation arising from strong cross-mode coherence. \textbf{d}, Superfluid regime at 100 GHz repetition rate, showing dense mode spacing and coherent interference fringes while maintaining high total on-chip comb power.}\label{Fig4}
\end{figure*}

The CV landscape of the Mott-insulator to superfluid transition in the pump-BG PhCR (Fig.~\ref{Fig3}d) evolves smoothly with $\alpha$ and $\gamma$; here, we additionally emphasize the role of the pump strength $f$. We compute CV from the backward-propagating spectrum $|b_\mu|^2$, since forward emission of the soliton is weak, and we evaluate CV over $\mu_r$, which varies slightly across the explored parameter space of $\alpha$ and $\gamma$. At fixed pump strength $f=2.8$, the largest values of $\gamma$ accessible in the CV landscape lead to breathing oscillations, which limit the experimentally reachable extent of the Mott-insulator regime. Therefore, the inset of Fig.~\ref{Fig3}d shows a CV map at a higher pump strength, $f=4.2$, illustrating that increasing $f$ shifts the accessible region in $(\alpha,\gamma)$ space. In particular, higher $f$ stabilizes soliton states at larger $\alpha$ and yields lower CV values, indicating that stronger pumping overall enhances self-interaction relative to cross-mode coupling and drives the system further toward the Mott-insulator regime.

To explore the CV landscape (Fig. \ref{Fig3}d (i)--(iv)), we perform experiments with 200 GHz PhCRs and pump the BG mode to form a soliton microcomb. Overall, the spectrum is smooth, since the hard edge of the lattice BG is missing and all the bosonic modes are connected through the Kerr nonlinearity, highlighting the critical role of $|b_0|^2$. First, we map the superfluid regime with relatively large $1/\gamma$. For the spectra in Fig.~\ref{Fig3}d(i,ii), we operate devices with $d_2=-0.052$ at a fixed pump strength $f=2.8$ while varying $\gamma$. At a small $\gamma=3.6$ (Fig.~\ref{Fig3}d(i)), the comb exhibits pronounced mode-to-mode non-uniformity, with a maximum power difference of $28~\mathrm{dB}$ and an experimentally extracted $\mathrm{CV}\approx 1.7$. Increasing $\gamma$ to 6.5 [Fig.~\ref{Fig3}d(ii)] enhances the comb near the edges of $\mu_r$, producing a visibly flatter envelope with a maximal power difference to $12~\mathrm{dB}$ and lowering the CV to $\approx 0.7$.

Guided by our theoretical CV mapping, to reach the Mott insulator we adjust the parameter settings of the PhCR so that the existence window of the target state is larger. We fabricate a device with $\gamma=10.9$ and larger normal dispersion ($d_2=-0.12$), and we operate it at a higher pump of $f=4.2$. As suggested by the CV map, this operating point lies in the Mott-insulator region, and soliton microcomb spectrum we measure with this device (Fig.~\ref{Fig3}d(iii)) reaches $\mathrm{CV}\approx 0.63$ with a maximal power difference of $8~\mathrm{dB}$ when evaluated over $\mu_r=\{\pm1,\pm2,\ldots,\pm15\}$. In this case, the non-uniformity is dominated by the boundary modes of $\mu_r$, consistent with dispersive-wave growth. Excluding the outermost three pairs and evaluating over $\mu_r=\{\pm1,\pm2,\ldots,\pm12\}$ yields $\mathrm{CV}=0.28$ and a reduced maximal mode difference of $3.5~\mathrm{dB}$, indicating the flattop Mott-insulator profile away from the dispersive waves.

Access to the Mott insulator to superfluid regime in pump-BG PhCRs enables us to explore very high power soliton microcombs, which are especially important for emerging applications. Figure \ref{Fig4} demonstrates soliton microcombs with $>$50 mW average comb power at dense repetition frequency of 100 GHz and 200 GHz. Moreover, all the PhCR soliton microcombs exhibit a smooth start-up behavior, namely we scan the pump laser frequency adiabatically onto resonance and stop upon comb formation. Figure~\ref{Fig4}a explores this behavior via LLE simulation as a function of detuning $\alpha$: as $\alpha$ is increased, a pair of primary sidebands emerges and seeds optical parametric oscillation (OPO), followed by the formation of a stable soliton state. 

Consistent with the motivation for high power, high coherence, and single-device generation across technologically relevant bands such as the telecom C band (191–196~THz), we demonstrate high per-mode power soliton microcombs in both the Mott-insulator and superfluid regimes. In the former shown by Fig.~\ref{Fig4}b, the total on-chip comb power (excluding the pump) reaches $54~\mathrm{mW}$. The comb lines have an on-chip power of more than -4~dBm for $\mu_r=\{\pm1, \ldots,\pm15\}$. 
With the outmost 2 pairs of modes that are dominated by dispersive-wave formation excluded, the CV for $\mu_r=\{\pm1,\pm2,\ldots,\pm13\}$ is 0.24 and the maximal mode difference is only 5.6 dB. 

In the superfluid regimes studied in Fig.~\ref{Fig4}c (200~GHz) and Fig.~\ref{Fig4}d (100~GHz), the total on-chip comb power reaches $92~\mathrm{mW}$ and $53~\mathrm{mW}$, respectively. The denser mode spacing exhibits a larger number of phase-matched four-wave mixing pathways to participate within the same spectral span. The resulting phase coherence across all the bosonic comb modes produces pronounced spectral interference fringes characteristic of the superfluid regime, reflecting strong cross-mode coupling and delocalized nonlinear dynamics. These interference structures arise from coherent multi-mode interactions rather than dispersion discontinuities, and their persistence at high power confirms that collective behavior remains robust in dense spacings. 

In summary, we demonstrate a non-equilibrium Mott-insulator to superfluid transition in driven–dissipative Kerr solitons, arising from universal competition between Kerr-mediated self-interaction and cross-mode coupling. By engineering either an extended lattice BG or a pump BG, we show that the collective state of a bosonic mode comb is deterministically reshaped, yielding controllable transitions between spectrally modulated and flattop soliton microcombs. Although the soliton microcomb operates at very large occupation, its dynamics are governed by the same Kerr interaction Hamiltonian that underlies correlated quantum matter. In this way, device geometry becomes a tool for Hamiltonian engineering, where deliberate customization of dispersion and BG coupling enables deterministic access to distinct collective states. The combination of high optical power and programmable many-body interactions opens opportunities in critical technological areas including millimeter wave synthesis, high capacity optical data links, and photonic AI acceleration.

We thank Grant Brodnik and Lindell Williams for reviewing the paper at NIST.  This research has been funded by AFOSR FA9550-20-1-0004 Project Number 19RT1019, NSF Quantum Leap Challenge Institute Award OMA – 2016244, DARPA NaPSAC, and NIST. The authors declare no competing interests. This work is a contribution of the US Government and is not subject to US copyright. Mention of specific companies or trade names is for scientific communication only and does not constitute an endorsement by NIST.

\section*{Methods}

\subsection{Normalization}

Throughout this work we employ the standard dimensionless normalization of the LLE to clarify the relative scales of dispersion, detuning, nonlinearity, and bandgap coupling. Unless otherwise stated, frequency parameters are normalized to the cavity half-linewidth $\kappa/2$, and time is normalized to the photon lifetime $2/\kappa$. In this convention, the dimensional bandgap is written as $\Gamma_\mu = \gamma_\mu \kappa$, where $\gamma_\mu$ denotes the normalized half-bandgap; the second-order dispersion is expressed as $D_2 = d_2 \kappa/2$, and the integrated dispersion as $D_\mathrm{int} = d_\mathrm{int} \kappa/2$. Time is rescaled according to $t = 2\tau/\kappa$, yielding a dimensionless evolution variable $\tau$. This normalization renders the LLE parameters directly comparable in units of cavity linewidth, allowing the competition between Kerr nonlinearity, dispersion, detuning, and bandgap coupling to be expressed on equal footing.


\subsection{Tantala PhCR fabrication and devices}

Microresonators are fabricated on thermally oxidized silicon wafers using amorphous tantalum pentoxide (Ta$_2$O$_5$) or titania–tantala films deposited by ion-beam sputtering (IBS) at room temperature \cite{Jung:21}. IBS produces dense, crack-free films with low optical loss and low intrinsic stress while remaining compatible with temperature-sensitive substrates. In selected devices, titania (TiO$_2$) is co-sputtered with tantala to form a homogeneous amorphous mixture (26\% volume fraction TiO$_2$), reducing oxygen-vacancy-related absorption and improving the intrinsic quality factor $Q_i$, particularly in oxide-clad resonators \cite{carollo2025amorphousmetaloxidemixtures}. Film thickness is selected to achieve the desired dispersion and cladding configuration: 570~nm for air-clad devices and 625~nm for oxide-clad devices to compensate for the effective-index shift introduced by the top cladding.

Photonic-crystal resonators (PhCRs) are patterned by electron-beam lithography using high-resolution nanopatterned resist. After development, an Al$_2$O$_3$ hard mask is deposited and lifted off, and the pattern is transferred into the tantala film using fluorine-based reactive-ion etching. This process yields smooth sidewalls and nanometer-scale control of the ring-width (RW) modulation that defines the photonic bandgap. The RW modulation period is designed to be twice the pump mode number to selectively split the targeted resonance while preserving broadband dispersion for Kerr nonlinear dynamics.

Following etching, wafers are annealed at 500$^\circ$C to reduce defect-related absorption and increase $Q_i$, accompanied by a small blue shift in resonance frequency due to a reduced refractive index. For fully encapsulated devices, a 3~$\mu$m SiO$_2$ top cladding is deposited by inductively coupled plasma–enhanced chemical vapor deposition (IC-PECVD) at 300$^\circ$C, below the crystallization threshold of tantala. The conformal, void-free oxide preserves nanoscale photonic-crystal features and provides environmental stability and additional dispersion engineering flexibility. A subsequent 500$^\circ$C anneal further reduces cladding-related absorption, yielding microresonators with $Q_i$ up to $\sim 4.5\times10^6$.

Wafers are diced into individual chips, and integrated pulley couplers are defined lithographically with the resonators to enable efficient excitation of the fundamental TE$_0$ mode. Coupling strength is controlled through geometry to access critical or slight over-coupling for soliton generation. This fabrication flow enables precise control of dispersion, bandgap strength, and nonlinear interaction parameters, providing a reproducible platform for Kerr soliton generation across repetition rates from 100~GHz to 400~GHz.

For the designs of the resonators, we apply a pulley coupler to couple the light into and out from the resonator. At the ends of the waveguides, we apply angled facets. 


\subsection*{PhC bandgap calibration}
The effective bandgap coupling strength $\gamma$ is controlled by the amplitude of the ring-width (RW) modulation that defines the photonic crystal (PhC). We calibrate $\gamma$ by measuring the cold-cavity mode splitting of the targeted resonance in linear transmission. For small modulation amplitudes, the induced splitting $\Gamma$ scales approximately linearly with the geometric modulation depth $A_{\mathrm{PhC}}$, reflecting the proportional increase in coherent backscattering between counter-propagating (or hybridized) mode families. In the normalized LLE framework, the dimensionless half-bandgap parameter is given by $\gamma = \Gamma/\kappa$, where $\kappa$ is the cavity linewidth. By systematically varying $A_{\mathrm{PhC}}$ and extracting the corresponding resonance splitting, we establish a direct calibration between fabrication geometry and the Hamiltonian bandgap parameter $\gamma$, enabling deterministic control of the effective self-interaction strength in the mode lattice.

\subsection*{The quantized coupled-mode LLE}

The quantized coupled-mode LLE equation can be written as \cite{PhysRevA.93.033820}
\begin{equation}\label{eq:qcpLLE}
\begin{split}
    \dot{\hata}_\mu &= \frac{1}{i\hbar}[\hata_\mu, \hath_\mathrm{tot}] - \hata_\mu,\\
    \dot{\hatb}_\mu &= \frac{1}{i\hbar}[\hatb_\mu, \hath_\mathrm{tot}] - \hatb_\mu,    
\end{split}
\end{equation}
where $\dot{\hata}_\mu=\partial_\tau \hata_\mu$, $\dot{\hatb}_\mu=\partial_\tau \hatb_\mu$, and the total Hamiltonian $\hath_\mathrm{tot} = \hath_\mathrm{free} + \hath_\mathrm{pump} + \hath_\mathrm{Kerr} + \hath_\mathrm{BG}$. Each of the terms can be expressed as
\begin{equation}
    \begin{split}
        \hath_\mathrm{free} &= \hbar \sum_\mu \left(\alpha+\frac{1}{2}d_2\mu^2-2P_b\right)\hata_\mu^\dagger \hata_\mu \\&+ \left(\alpha+\frac{1}{2}d_2\mu^2-2P_a\right)\hatb_\mu^\dagger \hatb_\mu, \\
    \hath_\mathrm{pump} &= i\hbar f(\hata_0^\dagger-\hata_0), \\
    \hath_\mathrm{BG} &= -\hbar \sum_\mu (\gamma_\mu \hata_\mu^\dagger \hatb_\mu + \gamma_\mu^* \hatb_\mu^\dagger \hata_\mu), \\
    \hath_\mathrm{Kerr} &= -\frac{\hbar}{2} \sum_{m, n, p,q} \delta_{m+p, n+q}(\hata_n^\dagger \hata_q^\dagger \hata_m \hata_p + \hatb_n^\dagger \hatb_q^\dagger \hatb_m \hatb_p)
    \end{split},
\end{equation}
and involves the pump laser detuning $\alpha$, the  second-order dispersion coefficient $d_2$, the pump amplitude $f$, the half-BG $\gamma_\mu$, and $P_a$ and $P_b$ are the total power in the forward and backward direction, respectively \cite{Kondratiev2020}.

\subsection{Kerr Hamiltonian}

The Kerr Hamiltonian $\hath_\mathrm{Kerr}$ contributes to both the self interaction (self-phase modulation) and cross interaction (cross-phase modulation and four-wave mixing). Hence, we decompose it as 
\begin{equation}
    \hath_\mathrm{Kerr} = \hath_\mathrm{SPM} + \hath_\mathrm{CPM} + \hath_\mathrm{FWM},
\end{equation}
where the self-phase modulation term (SPM, containing only 1 mode) is
\begin{equation}
    \hath_\mathrm{SPM} = -\frac{1}{2} \hbar \sum_\mu \left[ (\hata_\mu^\dagger)^2 (\hata_\mu)^2 + (\hatb_\mu^\dagger)^2 (\hatb_\mu)^2\right]
\end{equation}
the cross-phase modulation term (CPM, containing 2 distinct modes) is
\begin{equation}
    \hath_\mathrm{CPM} = -2\hbar \sum_{\mu<\nu} \left[\hata_\mu^\dagger \hata_\nu^\dagger \hata_\mu \hata_\nu + \hatb_\mu^\dagger \hatb_\nu^\dagger \hatb_\mu \hatb_\nu\right],
\end{equation}
and the four-wave mixing term $\hath_\mathrm{FWM}$ gathers the rest terms of $\hath_{Kerr}$ and contains 3 or 4 distinct modes.

\subsection{Interplay of self and cross interactions in Kerr soliton microcombs}

In the original Bose-Hubbard model, the competition between the self interaction (on-site energy $U$) and the cross interaction (the tunneling matrix $J$) decides whether the bosons are in the Mott insulator state or the superfluid state. This inspires us to divide the $\hath_\mathrm{tot}$ into three: the self interaction Hamiltonian $\hath_\mathrm{self}$, the cross interaction Hamiltonian $\hath_\mathrm{cross}$ and the pump Hamiltonian $\hath_\mathrm{pump}$,
\begin{equation}
    \hathm{tot} = \hathm{self} + \hathm{cross} + \hathm{pump,}
\end{equation}
where
\begin{equation}
\begin{split}
    \hath_\mathrm{self} &= \hathm{free}  + \hathm{SPM} + \hathm{BG}\\
    &=\hbar \sum_\mu \Big[
\left(\alpha+\frac{1}{2}d_2\mu^2-2P_b\right)
  \hata_\mu^\dagger \hata_\mu \\
&+ \left(\alpha+\frac{1}{2}d_2\mu^2-2P_a\right)
  \hatb_\mu^\dagger \hatb_\mu
\Big] \\& -\frac{1}{2} \hbar \sum_\mu \left[ (\hata_\mu^\dagger)^2 (\hata_\mu)^2 + (\hatb_\mu^\dagger)^2 (\hatb_\mu)^2\right] \\&
-\hbar \sum_\mu (\gamma_\mu \hata_\mu^\dagger \hatb_\mu + \gamma_\mu^* \hatb_\mu^\dagger \hata_\mu),
\end{split}
\end{equation}
and
\begin{align}
    \begin{split}
        \hathm{cross} &= \hathm{CPM} + \hathm{FWM}=\hathm{Kerr}-\hathm{SPM}.
    \end{split}
\end{align}
The factors that affect the self interaction include the detuning, the self-phase modulation and the bandgap, and the factors that affect the cross interaction include cross-phase modulation and four-wave mixing. We can control the relative strength between the self and cross interaction by changing the detuning $\alpha$, the half-bandgap $\gamma_\mu$ and the dispersion coefficient $d_2$. In this paper, we focus on the effects of $\alpha$ and $\gamma_\mu$.

\subsection{Mean field coupled-mode LLE equation}

The quantum operators $\hata_\mu$ and $\hatb_\mu$ obey
\begin{align}
\begin{split}
    [\hata_\mu, \hata_{\mu'}^\dagger] &= \delta_{\mu, \mu'}, \quad [\hatb_\mu, \hatb_{\mu'}^\dagger] = \delta_{\mu, \mu'}, \\ [\hata_\mu, \hata_{\mu'}]&=[\hatb_\mu, \hatb_{\mu'}]=[\hata_\mu^\dagger, \hata_{\mu'}^\dagger]=[\hatb_\mu^\dagger, \hatb_{\mu'}^\dagger]= 0,\\ [\hata_\mu, \hatb_{\mu'}] &=[\hata_\mu, \hatb_{\mu'}^\dagger]=[\hata_\mu^\dagger, \hatb_{\mu'}^\dagger] =[\hata_\mu^\dagger, \hatb_{\mu'}] = 0.
\end{split}
\end{align}

For the coherent states, $\lr{\hata_\mu}=a_\mu$, and $\lr{\hatb_\mu}=b_\mu$. Thus, the mean field expression of Eq. \ref{eq:qcpLLE} can be written as \cite{Kondratiev2020}
\begin{align}\label{eq:cpLLE}
\partial_\tau a_\mu
&= -(1+i\alpha)a_\mu
   - i\frac{d_2}{2}\mu^2 a_\mu
   + i\gamma_\mu b_\mu  \notag \\
&\quad + i \mathcal{F}(|a|^2a)_\mu
   + 2iP_b a_\mu
   + f\delta_{\mu,0},
\notag \\
\partial_\tau b_\mu
&= -(1+i\alpha)b_\mu
   - i\frac{d_2}{2}\mu^2 b_\mu
   + i\gamma_\mu^* a_\mu \notag \\
&\quad + i \mathcal{F}(|b|^2b)_\mu
   + 2iP_a b_\mu.
\end{align}
In the mean field regime, we apply Eq. \ref{eq:cpLLE} for simulation. 

\subsection{Definition of $\mu_r$}

In the pump-bandgap regime, only the pump mode is split. Therefore, we need to decide the bandwidth of the microcombs for CV calculation, i.e. $\mu_r$. Here we define the modes $\mu_r$ as the un-pumped modes ($\mu\neq 0$) on which $\langle\hathm{self}\rangle$ acts positively,  and considering $\lr{\hata_\mu^\dagger\hata_\mu}\ll \lr{\hatb_\mu^\dagger\hatb_\mu}$, we have
\begin{equation} \label{mu_r}
    \alpha+\frac{1}{2}d_2\mu^2-2P_a-\frac{1}{2}|b_\mu|^2>0, \quad \text{for } \mu\in \mu_r.
\end{equation}
Since $2P_a$ and $\frac{1}{2}|b_\mu|^2$ is small compared to $\alpha$, a good approximation for $\mu_r$ is $\{\pm1, \pm2, ... \pm \lfloor\sqrt{\frac{-2\alpha}{d2}}-1\rfloor \}$, where $\lfloor \cdot \rfloor$ is the floor function. In this paper, $\mu_r$ is calculated using Eq. \ref{mu_r} without approximation.

\subsection{Calculation of CV}

In simulation, we scan $\alpha$ from 0 to a large number for each $\gamma$, and then sweep back to get all the non-trivial solutions of Eq. \ref{eq:cpLLE}. We derive the CV map using the back-sweeping results. We calculate $|\psi_\mu|^2=|a_\mu|^2 + |b_\mu|^2$ in simulation, while in the experiments, we calculate the CV for the forward and backward fields ($|a_\mu|^2$ and $|b_\mu|^2$) separately. For the pump-bandgap regime, since $|a_\mu|^2\ll |b_\mu|^2$, the expression of CV is effectively
\begin{equation}
    \mathrm{CV} = \frac{\mathrm{Std}_{\mu \in \mu_r}(|b_\mu|^2)}{\mathrm{Mean}_{\mu \in \mu_r}(|b_\mu|^2)}.
\end{equation}

\bibliographystyle{apsrev4-2}
\bibliography{ref,sbib}

\end{document}